\documentstyle[aps,twocolumn,epsfig]{revtex}
\input epsf
\begin{document}
 \title{ On Density-Matrix Spectra for Two-Dimensional Quantum
  Systems}
 \author{
 Ming-Chiang Chung
 \, and Ingo Peschel \\
{\small Fachbereich Physik, Freie Universit\"at Berlin,} \\
{\small Arnimallee 14, D-14195 Berlin, Germany}}
 \maketitle
 \begin{abstract}
  For a two-dimensional system of coupled oscillators,
  the spectra of reduced density matrices can be obtained
  analytically. This provides an example where the
  features of these quantities, which are of central
  importance in numerical studies using the DMRG method,
  can be seen.
 \end{abstract}

  The density-matrix renormalization group method (DMRG)\cite{white}
  has brought an enormous progress for the study
  of one-dimensional quantum systems and
  related classical problems \cite{DMRG}. Consequently, this numerical
  approach has also been applied to  two-dimensional 
  quantum problems \cite{whitedjs,xiang}. 
  The situation there, however, appears to be much less favourable.
   
  The reason for this has to be sought in the properties 
  of the density matrices which are used
  to select an optimal reduced basis in the Hilbert space.
  The essential quantity is the distribution of their eigenvalues.
   In one dimension, one usually finds a rapid 
  decay, so that a relatively small number of states is 
  sufficient to give very good results. This basically
  exponential decay can be derived explicitely for non-critical 
  integrable models \cite{peschk,peschc}. In non-integrable cases,
  the spectra are less regular but still have similar 
  features \cite{oku,ritter}.
  
  The situation in two dimensions has been discussed in some
  detail for free fermions \cite{liang} and for the transverse Ising 
  model \cite{croo}. 
  It was found that, if one couples one-dimensional
  chains to form ladders, the number $m$ of states one needs
  to maintain a certain accuracy, grows exponentially with
  the width $M$ of the system. This was derived either from
  the limit of non-interacting chains, or from numerical
  calculations. The spectra themselves, however,  have not been 
  discussed so far, although they are at the core of the
  problem. It therefore seems worth while to treat an example,
  where one can give explicit results.
  
  This is possible for a system of coupled harmonic 
  oscillators, which is integrable in any number of
  dimensions. This problem was studied recently for the
  case of a linear chain \cite{peschc} and it was shown that the
  ground-state density matrices, either for one site or 
  for half of the system, are exponentials of bosonic 
  operators. This is a consequence of the Gaussian form of
  the ground state and holds quite generally. The problem 
  is only to determine the bosonic eigenvalues. This can
  be done either numerically for a small system, or 
  analytically in the thermodynamic limit.
  
  To be specific, consider the system described by the Hamiltonian
    \begin{eqnarray}
    H\;&=&\; \sum_{i}\,(-\frac{1}{2}\frac{\partial^2}{\partial u_i^2} 
    + \frac{1}{2}\,\omega_0^2\, u_i^2\,) \nonumber \\
    & &+  \sum_{i,j}\,
    \frac{1}{2}\, k_{ij} \,(u_{i}-u_{j})^2 \label{eqn:H}
    \end{eqnarray}
  where $u_{i}$ is the coordinate of the i-th oscillator and $\omega_{0}$ 
  its frequency. The masses are all equal to unity and the
  oscillators are coupled by springs of strenght $k_{ij}$. 
  Transforming $H$ to normal coordinates,
  one can write down the ground state immediately. In terms of
  the original coordinates it has the form
  \begin{equation}
      \phi\; =\; \exp(-\frac{1}{2}\sum_{i,j}\,A_{ij}\,u_{i}\,u_{j})         
  \label{eqn:phi} 
  \end{equation}    
  The total density matrix is then $|\phi><\phi|$. By integrating out
  part of the coordinates, one obtains reduced density matrices
  which have the diagonal form 
  \begin{equation}
      \rho\; =\;C\;
     \exp{(-\sum_{j}\,\varepsilon_{j}b^{\dagger}_{j}b_{j})}
     \label{eqn:rho}
  \end{equation}           
  with bosonic operators $b_j$,$b_{j}^{\dagger}$.
   The eigenvalues $\varepsilon_j$ follow from a
  matrix which is obtained from $A_{ij}$ and has a dimension equal to the 
  number of kept sites. One divides $A_{ij}$ into four submatrices 
   $a^{11},a^{12},a^{21},a^{22}$, according to whether the sites 
   $i$ and $j$ are kept or not. Then the matrix 
   $a^{11}[a^{12}(a^{22})^{-1}a^{21}]^{-1}$ has eigenvalues 
   $\cosh^2(\varepsilon_j/2)$.
  This can be shown by a straightforward generalization 
  of the approach in section 2 of \cite{peschc}. In this way,
  density-matrix spectra can be calculated numerically for an
  arbitrary assembly of coupled oscillators.

  For a large system, however, the situation simplifies. In \cite{peschc} it
  was shown that, for a chain with nearest-neighbour coupling $k$ and
  oscillator frequency $\omega_{0} = 1-k$, 
  the $\varepsilon_{j}$ for half of the system
  are, in the thermodynamic limit, given by 
  \begin{equation} 
      \varepsilon_{j}\; =\; (2j-1)\varepsilon, \;~~~~~~ j=1,2,3,\ldots               
      \label{eqn:epsilon}
  \end{equation} 
   where 
   \begin{equation}
     \varepsilon\; =\; \pi\,I(k')/I(k)
      \label{eqn:elliptic} 
   \end{equation}      
   Here $I(k)$ is the complete elliptic integral of the first kind
   and $k'= \sqrt{1-k^{2}}$. The result is also valid for finite systems
   if the size  is large compared with the correlation length. 
   The $\varepsilon_{j}$
   for smaller systems are still similar, but there are deviations
   from (\ref{eqn:epsilon}) which increase for larger values of $j$.

   Now consider a two-dimensional square lattice of oscillators with
   nearest-neighbour couplings $k_x$ and $k_y$ in the two directions. This 
   can be reduced to a one-dimensional problem by first introducing
   normal coordinates in the columns. The corresponding normal
   frequencies are
   \begin{equation}
     \omega(q)^2 = \omega_{0}^{2} + 2\, k_y\,(1-\cos{q})  
    \label{eqn:wq}                         
   \end{equation}    
   where the vertical momenta for open boundary conditions at the 
    ends and $M$ sites are given by
   \begin{equation}
          q\; =\;\frac{m}{M}\pi ,\;~~~~~~ m = 0,1,2,\ldots,(M-1)
    \label{eqn:pi}                 
   \end{equation}

    If one now couples the columns, the different momenta do not mix,
    and for each value of $q$ a horizontal chain of the form (\ref{eqn:H}) 
     results where the oscillator frequency is now $\omega(q)$ 
   and the coupling $k_x$. For the density matrix of the half-system,
    this leads to the spectrum (\ref{eqn:epsilon}),(\ref{eqn:elliptic})
    with the parameter $k = k(q)$
    determined from the relation $k_x/\omega(q) = k/(1-k)$ or explicitely
    \begin{equation}
               k = k_x/(k_x+\omega(q))\label{eqn:kw}
    \end{equation}                             
               
    In this way, an analytic expression for the spectrum is obtained.
    For each $j$, one has a band of $M$ eigenvalues $\varepsilon_j (q)$ due to 
    the transverse extension of the system. This reflects the 
    corresponding interface between the two parts into 
    which the system is divided.
    The dispersion of the vibrational modes in the
    vertical direction also determines the dispersion of the $\varepsilon$-band
    via (\ref{eqn:kw}). In particular, a large $\omega(q)$
    also leads to a large value of $\varepsilon$.

     \begin{figure}
     \begin{center}
     \epsfxsize=70mm
     \epsfysize=60mm
     \epsffile{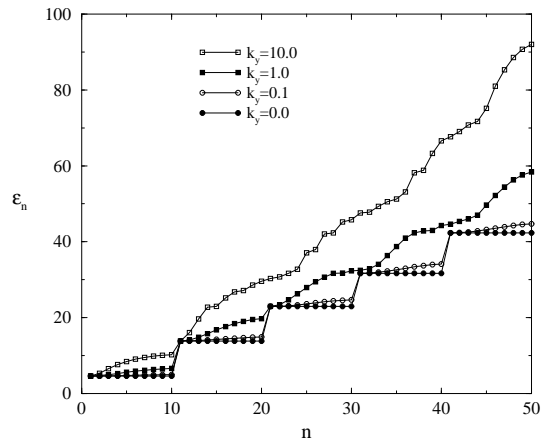}
     \caption{\label{fig1}  Bosonic single-particle
      eigenvalues $\varepsilon_j (q)$, c.f. Equ.(\ref{eqn:rho}),
    for half of a $10\times 10$ system, arranged in ascending order,
           for $\omega_0 = k_x = 1.0$ and four values of the coupling $k_y$. 
    }
     \end{center}
     \end{figure}

    Such spectra, calculated numerically for a $10\times10$ lattice,
    are show in Fig.\ref{fig1}. Plotted are the 
    $\varepsilon_j (q)$ arranged according
    to their magnitude for different values of the transverse
    coupling $k_y$. For non-interacting horizontal chains, one has
    a sequence of plateaus. Turning on $k_y (>0)$, the eigenvalues 
    increase except for $q = 0$ and form real bands. At the
    lower end, the stair-like structure still persists, while for
    larger values of $j$ the bands are spread more due to the factor 
    $(2j-1)$ and overlap eventually. After the proper ordering
    of the $\varepsilon$, a continuous curve emerges. It corresponds roughly
    to a linear relation of the form
    \begin{equation}
               \varepsilon_n\;\cong\; \lambda\,n                       
    \label{eqn:lambda}
    \end{equation}            
    with integer $n$ and $\lambda \cong 2\,\varepsilon(q=0)/M$
     inversely proportional to the width $M$.

\begin{figure}
     \begin{center}
     \epsfxsize=70mm
     \epsfysize=60mm
     \epsffile{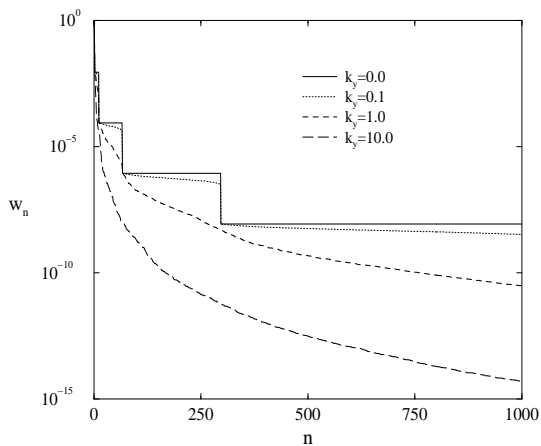}
     \caption{\label{fig2}  Density-matrix eigenvalues $w_n$,
       arranged in decreasing
       order, obtained from the $\varepsilon_j(q)$ in Fig.\ref{fig1} and for
           the same parameters
        }
     \end{center}
     \end{figure}

    The actual eigenvalues $w_n$ of $\rho$ are obtained by 
    specifying the occupation numbers of the bosonic single-particle levels
    $\varepsilon_j (q)$. This leads to an increasingly larger number of 
   possibilities as more $\varepsilon$ are involved, 
   i.e. for smaller $w_n$. The final 
    result is shown in Fig.\ref{fig2} for 
    the same parameters as in Fig.\ref{fig1}.
    One can see that the stair-like structure persists also in the
    $w_n$ for small $k_y$, although the plateaus are much longer and given
    by combinatorical factors. For larger $k_y$, rather smooth curves
    arise which drop increasingly faster. In all cases, there is
    a rapid initial decay followed by a slower decrease for larger $n$.
    Following \cite{oku}, one can derive an asymptotic formula from
    (\ref{eqn:lambda}) which reads
   
    \begin{equation}
     w_n \;\sim\; \exp\{-(\lambda/(2\,\pi^2/3))\ln^{2} n\}
    \label{eqn:wn}
    \end{equation}    
    and which is obeyed reasonably well by the curves. Due to the slow 
    decay, the truncation error when cutting off the spectrum also
    decreases slowly. After $n =100$, $500$ and $1000$ it is approximately
    $10^{-5}$, $10^{-7}$ and $10^{-8}$, 
    respectively, if $k_x=k_y=\omega_0=1.0$.

\begin{figure}
     \begin{center}
     \epsfxsize=70mm
     \epsfysize=60mm
     \epsffile{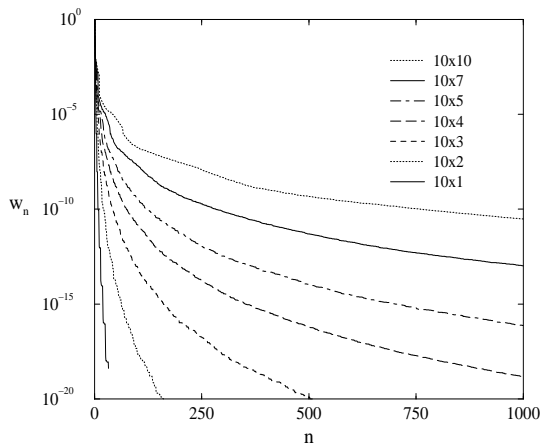}
     \caption{\label{fig3}  Density-matrix eigenvalues $w_n$ 
        for systems of different
           width $M$ and $\omega_0 = k_x = k_y = 1.0$
      }
     \end{center}
     \end{figure}

    The dependence of the $w_n$-spectrum on 
    the width $M$ is shown in Fig.\ref{fig3}
    for the case $k_x = k_y = 1$. One can see how the curves drop more and
    more slowly as $M$ increases, in accordance with (\ref{eqn:lambda})
     and (\ref{eqn:wn}) and
    the decrease of $\lambda$ with $M$.

    These results confirm that the situation worsens as the
    system becomes more two-dimensional. The faster initial 
    decay before the onset of the combinatorical effects helps
    in numerical calculations. Also the interaction helps here
     to some extent since the $\varepsilon$-values increase with $k_y$
    , but this does not remove the basic $1/M$ dependence
    in the exponent. The same features are found if one
    assumes that one is dealing with fermionic operators
    in (\ref{eqn:rho}). This would correspond to a fermionic system
    with pair terms such that the Hamiltonian expressed in Fermi operators 
    has the same structure 
    as (\ref{eqn:H}) expressed in Bose operators.
    In this case, the combinatorical possibilities are
    reduced, but this leads only to a change $\lambda \to 2\lambda$ in
    (\ref{eqn:wn}).  One should also note that we have treated a non-critical
    system where the situation is in general more favourable.
    One could extend the considerations to three dimensions,
    in which case one has two momenta for the transverse 
    directions and therefore an even larger number of
    $\varepsilon$-values in each band.

\begin{figure}
     \begin{center}
     \epsfxsize=70mm
     \epsfysize=60mm
     \epsffile{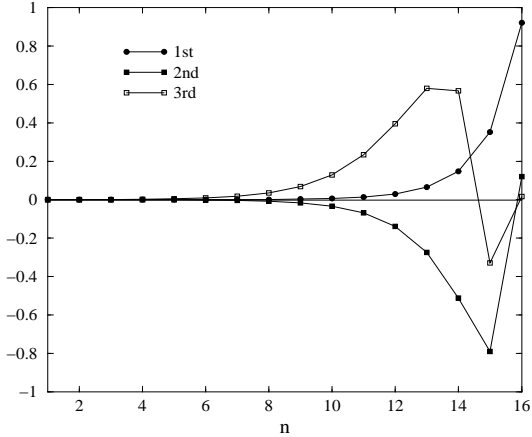}
     \caption{\label{fig4}  Density-matrix eigenstates for the left part 
      of a chain of 32 
      sites. Shown are the amplitudes as a function of the position 
      for the lowest three $\varepsilon_j$-values for $\omega_0 = k = 0.5$.
      }
     \end{center}
     \end{figure}

    Returning briefly to one dimension, we would like to mention that 
    our model also shows the origin of the different DMRG performance
    for chains and rings \cite{white} very clearly. This problem is,
    in fact, closely related to those discussed above. If one calculates
     the density-matrix spectrum for a half-ring, one finds the same 
    $\varepsilon_j$-values at the lower end as for the half-chain,
    but each value appears {\it twice}. When plotted, this leads to
    a structure as in Fig.\ref{fig1} with steps of length two. 
    The reason lies in the form of the eigenstates of $\rho$
    which, for small $\varepsilon_j$, are concentrated near the boundary
    between system and environment. This feature, which was conjectured
    before \cite{white,huse}, can be seen explicitely here and 
    is illustrated in Fig.\ref{fig4}. The effect is known from the 
    closely related corner transfer matrix of the massive Gaussian 
    model \cite{truong}. For a half-ring, which has two points 
    of contact, one then finds two such sets of states which are 
    approximately independent of each other for small $\varepsilon_j$.
    Therefore, $\rho \cong \rho_{L}\cdot \rho_{R}$ where $\rho_{L}$ 
    and $\rho_{R}$ are density matrices for only a left or only a right
    bondary. Thus the situation is the same as for a ladder consisting 
    of two only weakly interacting chains.

    Coming back to the two-dimensional problem, the spectra found here 
    explain, in a very direct way, the difficulties of the DMRG in this case.
    To apply the method, one should use as many symmetries as possible 
    \cite{croo}. However, to treat really large systems, 
   a procedure which avoids the extended interfaces between the parts of 
   the system would be necessary. Whether the momentum-space approach
    of Ref.\cite{xiang} can help here, is not yet clear.     
 
  \vspace{0.5cm}

  \noindent{\normalsize\bf Acknowledgements} 
  
  \vspace{0.2cm}
   We thank X. Wang for discussions and A.~Gendiar 
   for correspondence. M.C.~Chung acknowledges
   the support of Deutscher Akademischer Austauschdienst (DAAD).

 \end{document}